\documentclass{aa}

\usepackage[varg]{txfonts}
\usepackage[colorlinks=true, linkcolor=blue, citecolor=blue, urlcolor=gray]{hyperref}
\usepackage{xcolor}

\begin{document}

\title{A new analytical scattering phase function for interstellar dust}

\author{Maarten Baes, Peter Camps, and Anand Utsav Kapoor}
\authorrunning{M.\ Baes et al.}

\institute{Sterrenkundig Observatorium, Universiteit Gent, Krijgslaan 281 S9, B-9000 Gent, Belgium}

\date{\today} 

\abstract{Properly modelling scattering by interstellar dust grains requires a good characterisation of the scattering phase function. The Henyey-Greenstein phase function has become the standard for describing anisotropic scattering by dust grains, but it is a poor representation of the real scattering phase function outside the optical range.}
{We investigate alternatives for the Henyey-Greenstein phase function that would allow the scattering properties of dust grains to be described. Our goal is to find a balance between realism and complexity: the scattering phase function should be flexible enough to provide an accurate fit to the scattering properties of dust grains over a wide wavelength range, and it should be simple enough to be easy to handle, especially in the context of radiative transfer calculations.}
{We fit various analytical phase functions to the scattering phase function corresponding to the BARE-GR-S model, one of the most popular and commonly adopted models for interstellar dust. We weigh the accuracy of the fit against the number of free parameters in the analytical phase functions.}
{We confirm that the Henyey-Greenstein phase functions poorly describe scattering by dust grains, particularly at ultraviolet (UV) wavelengths, with relative differences of up to 50\%. The Draine phase function alleviates this problem at near-infrared (NIR) wavelengths, but not in the UV. The two-term Reynolds-McCormick phase function, recently advocated in the context of light scattering in nanoscale materials and aquatic media, describes the BARE-GR-S data very well, but its five free parameters are degenerate. We propose a simpler phase function, the two-term ultraspherical-2 (TTU2) phase function, that also provides an excellent fit to the BARE-GR-S phase function over the entire UV-NIR wavelength range. This new phase function is characterised by three free parameters with a simple physical interpretation. We demonstrate that the TTU2 phase function is easily integrated in both the spherical harmonics and the Monte Carlo radiative transfer approaches, without a significant overhead or increased complexity.}
{The new TTU2 phase function provides an ideal balance between being simple enough to be easily adopted and realistic enough to accurately describe scattering by dust grains. We advocate its application in astrophysical applications, in particular in dust radiative transfer calculations.}


\maketitle

\section{Introduction}

Cosmic dust is a vital ingredient of the circumstellar and interstellar media. Dust grains have an important role in several astrophysical processes, including the photoelectric heating of interstellar gas \citep{1978ApJS...36..595D} and the formation of molecular hydrogen \citep{1963ApJ...138..393G, 1971ApJ...163..155H, 2014A&A...569A.100B}. One of the most important characteristics of cosmic dust grains is the efficient interaction with radiation: dust grains efficiently absorb ultraviolet (UV) and optical radiation and subsequently re-emit the absorbed energy as thermal radiation in the infrared. Even though the dust-to-stellar mass ratio in galaxies typically ranges between $10^{-5}$ and $10^{-3}$ \citep{2019A&A...624A..80N}, on average one-third of all starlight emitted in galaxies is absorbed by dust and transformed to thermal infrared radiation \citep{2002MNRAS.335L..41P, 2016A&A...586A..13V, 2018A&A...620A.112B}. 

Dust grains are also very efficient at scattering radiation, especially at UV and optical wavelengths. At these wavelengths, absorption and scattering have roughly the same importance \citep{2003ARA&A..41..241D, 2004ApJS..152..211Z}. This implies that if we want to infer the intrinsic properties of cosmic systems from UV and optical radiation, we need to properly account for scattering effects. Properly modelling scattering is particularly challenging as multiple scattering effects are often hard to predict or are counter-intuitive \citep[see e.g.][]{1992ApJ...393..611W, 1994ApJ...432..114B}.

Properly modelling scattering by interstellar dust requires a good characterisation of the scattering phase function, that is, the probability density of the scattering angle after a single scattering event. Observations of reflection nebulae, dark clouds, and diffuse Galactic light have shown that scattering by interstellar grains is strongly anisotropic, especially at short wavelengths \citep{1976ApJ...208...64L, 1992ApJ...395L...5W, 1995ApJ...446L..97C, 2003ApJ...589..362G}. 

The Henyey-Greenstein (HG) phase function was originally introduced by \citet{1941ApJ....93...70H} to describe scattering by interstellar dust grains in the Milky Way. It has become the standard phase function for describing anisotropic scattering by dust grains in astronomy \citep{1988ApJ...333..673B, 1996ApJ...463..681W, 1999A&A...344..868X, 1999ApJS..123..437F, 2012MNRAS.419.1913B, 2014MNRAS.441..869D} and has been widely used in other fields as well, including atmospheric sciences \citep{1969JAtS...26.1078D, Boucher1998}, oceanography \citep{Hakim:03, 2004PrOce..61...27S, doi:10.1029/2002JC001513}, and biomedical imaging \citep{10.1117/12.148348, 1999InvPr..15R..41A, 1999JBO.....4...36R}. The HG phase function is a simple analytical formula that depends on just a single parameter, $g$. By varying this anisotropy parameter, the HG phase function can describe a wide variety of scattering regimes, from completely isotropic to extremely forward or backward scattering. Another interesting characteristic of the HG phase function is that it is easily integrated into the most popular radiative transfer solution methods. Unfortunately, the HG phase function has a number of disadvantages too. The main disadvantage is that it is a relatively poor approximation to the real phase function for scattering off dust grains outside the optical range, in particular at UV and near-infrared (NIR) wavelengths \citep{2003ApJ...598.1017D, 2008ApJS..177..546S}.

Several alternatives have been proposed for the HG phase function. One interesting alternative was proposed by \citet{2003ApJ...598.1017D}, a two-parameter generalisation of the HG phase function. It has a number of interesting characteristics and has been shown to provide a better approximation to the scattering function for interstellar dust at optical and NIR wavelengths compared to the classic HG phase function. At UV wavelengths, however, the Draine phase function still poorly describes the scattering properties of astrophysical dust. An alternative two-parameter extension of the HG phase function was presented by \citet{RM80} and is known as the Reynolds-McCormick (RM) or the Gegenbauer kernel phase function. The advantage of the RM over the classic HG phase function is that it better describes strongly anisotropic scattering. It has primarily been applied in the context of scattering in biological tissues and fluids \citep[e.g.][]{10.1117/12.273654, Hammer98, Lindbergh09, Karlsson:12, Calabro15}. The RM phase function offers more flexibility than the one-parameter HG phase function but generally fails for distributions that are both forward and backward peaked. This is particularly problematic in the UV and NIR regimes, where the scattering phase function corresponding to interstellar dust grains shows this feature.  A solution to that problem can be to use a linear combination of two HG scattering functions, resulting in the two-term Henyey-Greenstein (TTHG) phase function \citep{1965ApJ...142.1563I, 1968ApJ...152..823I}. 

\citet{C9NR01707K} recently presented an interesting study of the light scattering properties from a polymer layer containing sub-micron metallic and dielectric particles. They experimentally measured the scattering phase function, which exhibited strongly forward and backward peaked scattering patterns. They proposed a two-term Reynolds-McCormick (TTRM) phase function that is a linear combination of two RM phase functions, resulting in a five-parameter analytical phase function. They fitted this phase function and several other analytical phase functions to their experimental data and found that the TTRM phase function consistently gave the best fit in all cases. Very recently, \citet{Harmel2021} confirmed the superiority of the TTRM phase function proposed by \citet{C9NR01707K} to describe the scattering by microalgae and mineral hydrosols in water.

The main issue with this TTRM phase function is that it contains five free parameters and hence lacks the attractive simplicity and ease of use of a simpler function, such as the HG phase function. The same disadvantage applies to more numerical phase functions, which are often based on expansions of the phase function in terms of Legendre or Chebyshev polynomials \citep[e.g.][]{1950ratr.book.....C, 1998JQSRT..60.1001S, 2008ApJS..177..546S, Zhang2017}.

In this paper we investigate alternatives for the HG phase function to describe the scattering properties of dust grains in the interstellar medium in galaxies. Our goal is to find a balance between realism and complexity: on the one hand the scattering phase function should be flexible enough to provide an accurate fit to the actual scattering properties of dust grains over a wide wavelength range, and on the other hand it should be simple enough to be easy to handle, especially in the context of radiative transfer calculations. 

This paper is structured as follows. In Sect.~{\ref{DustModel.sec}} we present the dust model and the resulting empirical scattering phase function data we use for our calculations. In Sect.~{\ref{PhaseFunctions.sec}} we fit different analytical phase functions to the empirical data. In Sect.~{\ref{newpf.sec}} we present a new three-parameter phase function and discuss its properties and benefits. Our conclusions are summed up in Sect.~{\ref{Summary.sec}}.

\section{Dust model}
\label{DustModel.sec}

\begin{figure*}
\centering
\includegraphics[width=0.96\textwidth]{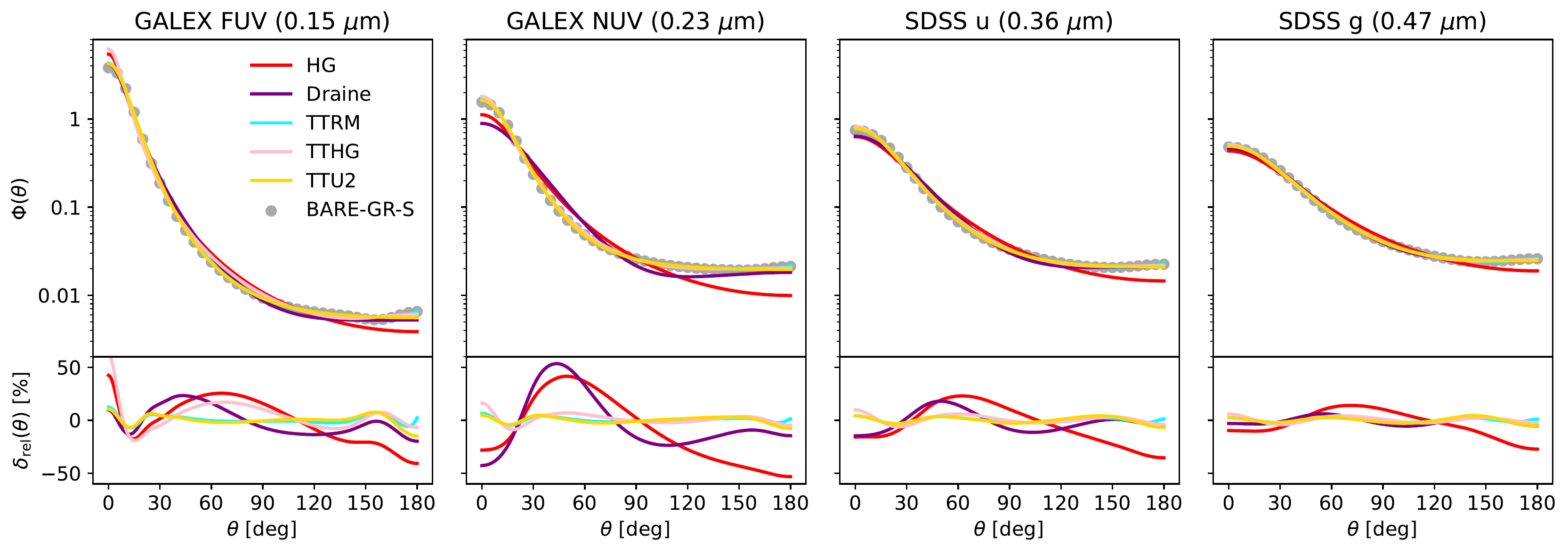}\hspace*{2em}\\
\includegraphics[width=0.96\textwidth]{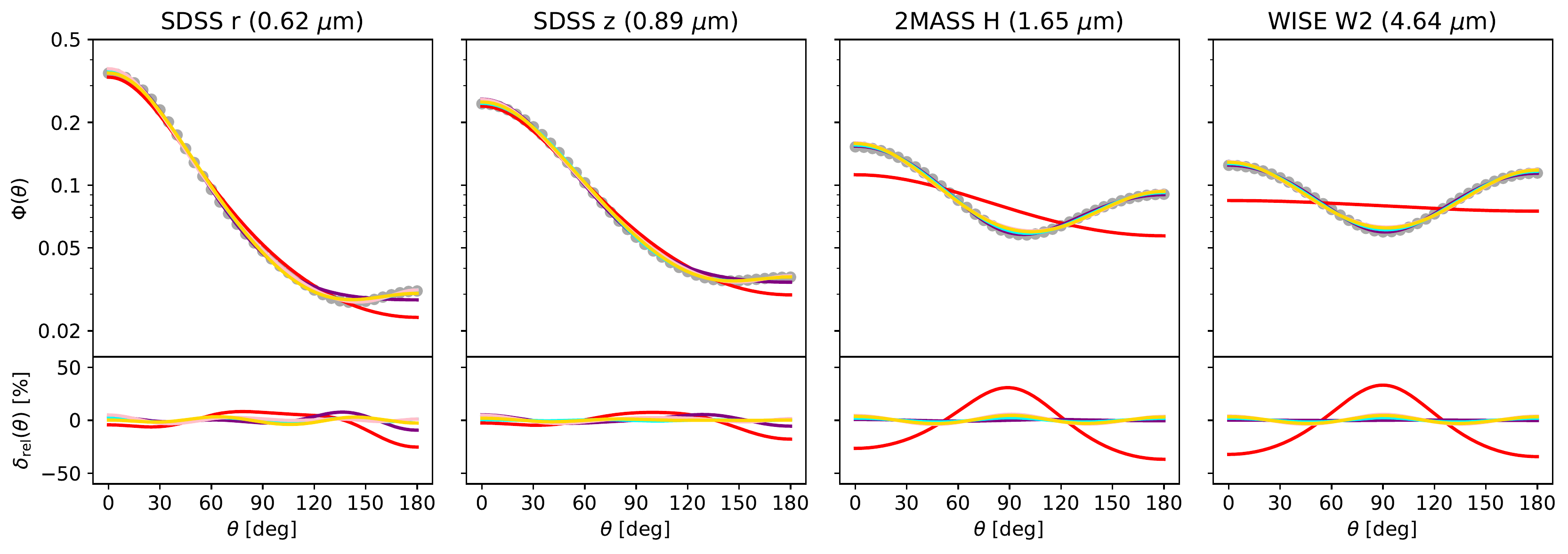}\hspace*{2em}
\caption{Empirical scattering phase functions (grey dots) of the \citet{2004ApJS..152..211Z} BARE-GR-S dust model at the effective wavelengths of eight common UV--NIR bands. For the sake of clarity we only plotted the data points at a resolution of 5 deg, whereas the BARE-GR-S dataset contains phase function data at 1 deg resolution. The solid lines in each upper panel correspond to analytical phase function fits to these empirical data: the HG phase function (red, Sect.~{\ref{HG.sec}}), the Draine phase function (purple, Sect.~{\ref{Draine.sec}}), the TTRM phase function (cyan, Sect.~{\ref{TTRM.sec}}), and the new TTU2 phase function (gold, Sect.~{\ref{newpf.sec}}). The bottom panels correspond to the relative difference between the analytical model and the empirical data.} 
\label{all-fits.fig}
\end{figure*}

\begin{figure*}
\centering
\includegraphics[width=0.96\textwidth]{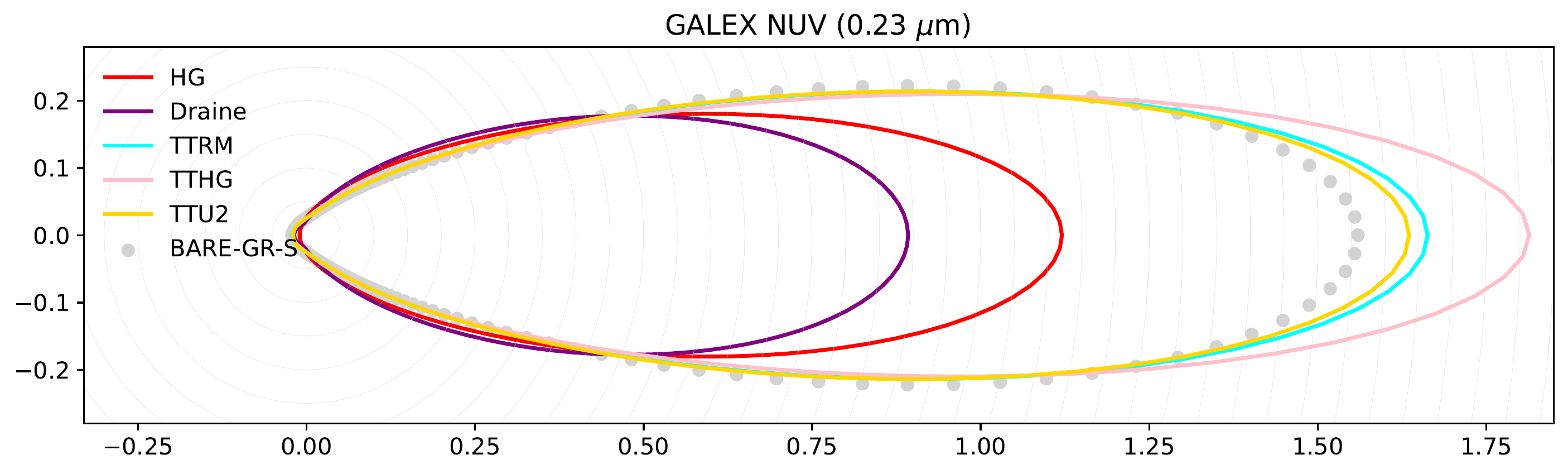}\hspace*{2em}\\
\includegraphics[width=0.505\textwidth]{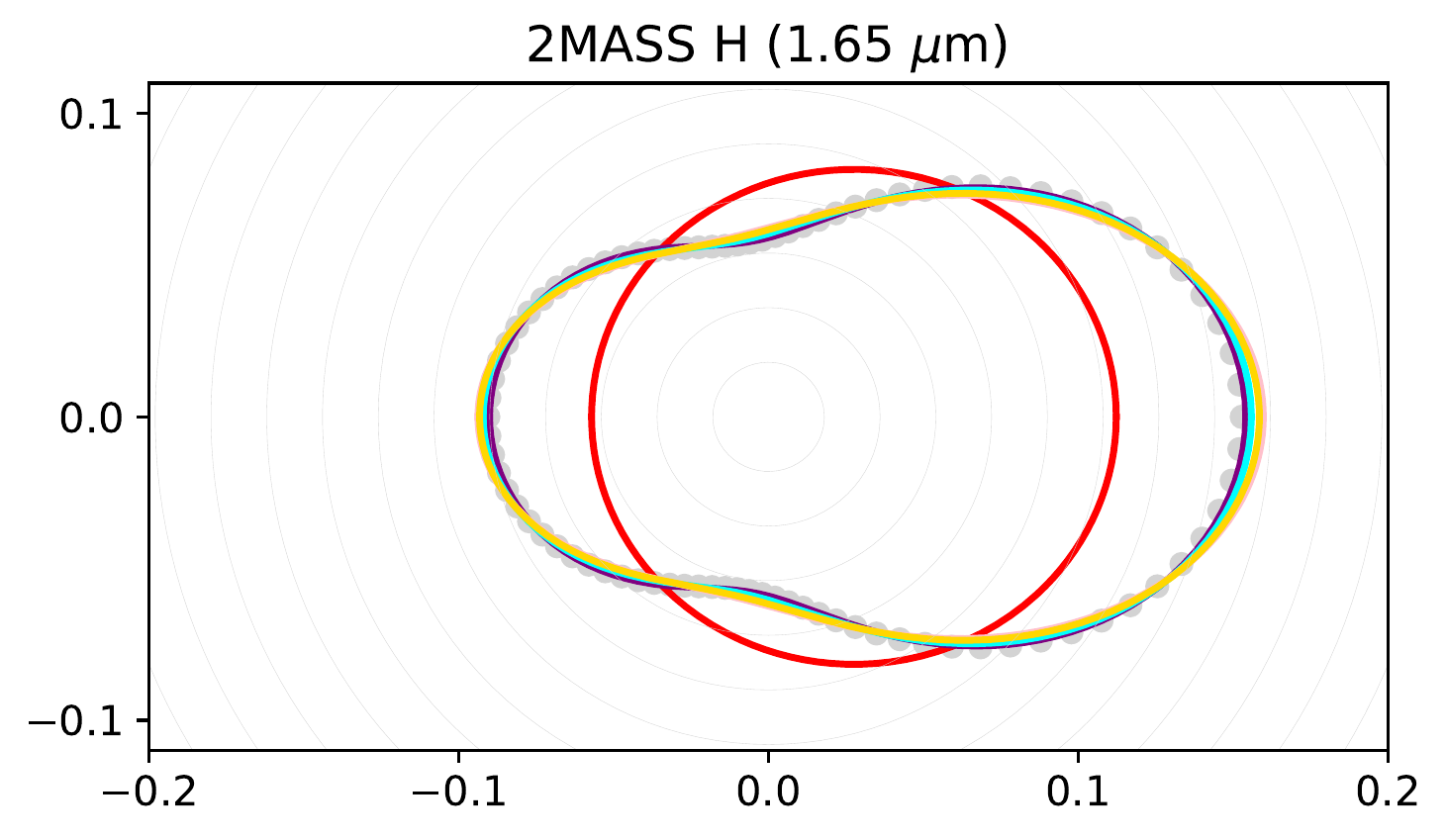}\hspace*{2em}
\caption{Scattering phase function for two representative wavelengths (0.23 and 1.65~$\mu$m), shown in polar format. The meanings of the lines and symbols are the same as in Fig.~{\ref{all-fits.fig}}.} 
\label{pp-polar.fig}
\end{figure*}

We adopted the BARE-GR-S dust model presented by \citet{2004ApJS..152..211Z}, one of the most popular and commonly adopted models for interstellar dust. It consists of a mixture of spherical graphite and silicate grains and polycyclic aromatic hydrocarbons (PAHs). The size distribution and relative abundances of each component are derived by simultaneously fitting the far-ultraviolet (FUV) to NIR extinction, the diffuse infrared emission in the Milky Way, and elemental abundance constraints on the dust. 

The BARE-GR-S model was used as the standard dust model in the suite of TRUST radiative transfer benchmark studies \citep{2015A&A...580A..87C, 2017A&A...603A.114G}. In the frame of this benchmarking effort, data files containing the detailed optical properties of the dust mixture are available online\footnote{\url{http://www.shg.ugent.be}}. In particular, size- and population-integrated effective scattering phase functions are available with a 1~degree angular resolution for 1201 wavelengths from 10~\AA\ to 10~mm. Throughout this paper we use the normalisation convention
\begin{equation}
\int \Phi(\theta)\,{\text{d}}\Omega = 1.
\label{norm}
\end{equation}
We denote these BARE-GR-S as the empirical phase functions. 

In this paper we focus on the UV--NIR wavelength regime, covering the range from about 0.1 to 5~$\mu$m. In Fig.~{\ref{all-fits.fig}} we show the empirical scattering phase function $\Phi(\theta)$ as a function of the scattering angle $\theta$ at the effective wavelengths of several standard broadband filters. In Fig.~{\ref{pp-polar.fig}} we show two representative cases in the more familiar polar representation. At UV wavelengths, the scattering is strongly forward peaked ($\theta\sim0^\circ$) with a very minor secondary backward peak ($\theta\sim180^\circ$). As the wavelength increases into the optical regime, the phase function becomes gradually less forward peaked (i.e. more isotropic), but the relative contribution of the backward scattering peak gradually increases. At NIR wavelengths, the phase function is largely isotropic, with a forward scattering peak that is only mildly dominant over the backward scattering peak. At the longest wavelengths considered here, the empirical phase function converges to the Rayleigh phase function,
\begin{equation}
\Phi_{\text{R}}(\theta) = \frac{3}{16\pi}\left(1+\cos^2\theta\right).
\label{pR}
\end{equation}

\section{Analytical phase functions}
\label{PhaseFunctions.sec}

In this section, we fit different analytical phase functions used in the literature to the empirical data corresponding to the BARE-GR-S dust model. For an analytical phase function $\Phi_{\text{ana}}(\theta,{\boldsymbol{p}})$ characterised by free parameters ${\boldsymbol{p}}$, we determine at each wavelength the best-fitting parameters ${\boldsymbol{p}}_{\text{fit}}$ by minimising the relative error, defined by \citet{2003ApJ...598.1017D} as
\begin{equation}
h_{\text{rel}}({\boldsymbol{p}}) = \left\{ \int \frac{{\text{d}}\Omega}{4\pi} \left[ \frac{\Phi(\theta)-\Phi_{\text{ana}}(\theta, {\boldsymbol{p}})}{\Phi(\theta)}\right]^2 \right\}^{1/2}.
\end{equation}
This metric is used to determine the best-fitting parameters, and can be used to compare the global quality of the fit of different analytical phase functions. For a more detailed investigation of how well a given parameterisation describes the empirical phase function, it is also useful to consider the relative difference between the empirical phase function and the best-fitting analytical phase function,
\begin{equation}
\delta_{\text{rel}}(\theta) = \frac{\Phi(\theta)-\Phi_{\text{ana}}(\theta, {\boldsymbol{p}}_{\text{fit}})}{\Phi(\theta)},
\end{equation}
explicitly as a function of the scattering angle $\theta$.

\subsection{The HG phase function}
\label{HG.sec}

Scattering off dust grains is usually described using the HG phase function, 
\begin{equation}
\Phi_{\text{HG}}(\theta,g) = \frac{1}{4\pi}\,\frac{1-g^2}{(1+g^2-2g\cos\theta)^{3/2}}.
\label{pHG}
\end{equation}
This expression contains a single free parameter, $g$, which in this particular case is nothing but the anisotropy parameter
\begin{equation}
g = \langle\cos\theta\rangle_{\text{HG}} \equiv \int \Phi_{\text{HG}}(\theta, g) \cos\theta\, {\text{d}}\Omega.
\end{equation}
In general, the anisotropy parameter indicates to which degree the scattering phase function deviates from isotropy. The wavelength dependence of the asymmetry parameter is usually one of the constraints or sanity checks for the construction of physical dust models \citep[e.g.][]{2004ApJS..152..211Z, 2014A&A...561A..82S, 2018A&A...617A.124Y, 2021ApJ...906...73H}. The fact that the HG phase function contains the anisotropy parameter as an explicit free parameter makes it very attractive, and explains to some degree its popularity. 

The solid red lines in Figs.~{\ref{all-fits.fig}} and~{\ref{pp-polar.fig}} represent HG phase functions fitted to the empirical BARE-GR-S data.\footnote{As discussed in the first paragraph of Sec.~{\ref{PhaseFunctions.sec}}, we have determined $g$ at each wavelength by minimising the relative error $h_{\text{rel}}(g)$. An alternative option could have been choosing $g$ equal to the numerically determined anisotropy parameter of the empirical phase function. Advantages of our approach are that it can be applied consistently to all analytical phase functions, and that, by definition, it selects the parameters corresponding the lowest relative error. The resulting $g$ values of both approaches are very similar.} Overall, the HG phase function provides a reasonable first-order approximation to the actual phase function, but there are significant deviations. These are most clearly seen in the bottom parts of each panel of Fig.~{\ref{all-fits.fig}}, where we show the relative difference between the empirical phase function and the best-fitting HG phase function. The HG phase function underestimates the backward scattering peak at all wavelengths. At UV and NIR wavelengths the underestimation amounts up to 30\%, while at optical wavelengths it is reduced to $\sim20$\%. Similarly, the HG phase function underestimates the forward scattering peak of the phase function at all wavelengths, with magnitudes varying from $\sim10$\% in the optical up to 30\% in the NIR and more than 40\% in the near-ultraviolet (NUV) band. The only exception is the FUV band, where the distribution function has a particular shape in the forward direction and the actual peak is underestimated by 30\%. For scattering angles around 60$^\circ$, the HG phase function overestimates the empirical phase function by different degrees, ranging from around 10\% in the optical up to 30\% in the NIR and even up to 50\% in the NUV.

\subsection{The Draine phase function}
\label{Draine.sec}

As indicated in the Introduction, a generalisation of the HG phase function was proposed by \citet{2003ApJ...598.1017D}. He proposed the following analytical phase function: 
\begin{equation}
\Phi_{\text{D}}(\theta, g, \alpha) = \frac{3}{4\pi}\, \frac{1-g^2}{3+\alpha(1+2g^2)}\, \frac{1+\alpha\cos^2\theta}{(1+g^2-2g\cos\theta)^{3/2}}.
\label{pD}
\end{equation}
This phase function was designed to bridge the cases of HG and Rayleigh scattering: for $\alpha=0$ the Draine phase function (\ref{pD}) reduces to the HG phase function~(\ref{pHG}), for $g=0$ and $\alpha=1$ it reduces to the Rayleigh phase function (\ref{pR}).

The purple lines in Figs.~{\ref{all-fits.fig}} and~{\ref{pp-polar.fig}} represent the best fitting Draine phase functions to the empirical BARE-GR-S data. Compared to the HG phase function, the improvement is particularly evident at the longest wavelengths, shown on the bottom row. This is not surprising as the Draine phase function is a generalisation of both the HG phase function and the Rayleigh phase function that is applicable at long wavelengths. At blue and particularly UV wavelengths there is less improvement. In the FUV the fit to the scattering phase function is slightly improved, but in the NUV band the forward scattering peak is still underestimated by about 40\% and the phase function at $\sim60^\circ$ is overestimated by 50\%.

\subsection{The TTHG phase function}
\label{Draine.sec}

The TTHG phase function consists of a linear combination of two HG phase functions, one with a forward scattering and one with a backward scattering peak,
\begin{equation}
\Phi_{\text{TTHG}}(\theta, g_1, g_2, \gamma) = \gamma\,\Phi_{\text{HG}}(\theta, g_1) 
+ (1-\gamma)\,\Phi_{\text{HG}}(\theta, g_2).
\label{pTTHG}
\end{equation}
This three-parameter phase function was first proposed by \citet{1965ApJ...142.1563I, 1968ApJ...152..823I} and subsequently applied by several other authors \citep[e.g.][]{1975JQSRT..15..839K, 1977ApJS...35....1W, 2016JQSRT.184...40Z, 2021Icar..35414021C}. 

The pink lines in Figs.~{\ref{all-fits.fig}} and~{\ref{pp-polar.fig}} represent the best fitting TTHG phase functions to the empirical BARE-GR-S data. Compared to the single HG phase function, the TTHG phase function results in a significant improvement in the fit quality over the entire wavelength range. In particular, by combining two HG phase functions with a similar relative weight and with $g_1\approx-g_2$, the phase function at NIR wavelengths can properly be fitted (see bottom panel of Fig.~{\ref{pp-polar.fig}}).  At UV wavelengths, a significant discrepancy remains, as clearly illustrated in the top panel of Fig.~{\ref{pp-polar.fig}}.

\subsection{The TTRM phase function}
\label{TTRM.sec}

Inspired by the conclusions by \citet{C9NR01707K} and \citet{Harmel2021}, our next step was to test whether the TTRM phase function provides a suitable option to describe the scattering phase function for interstellar dust. The TTRM phase function can be written as
\begin{multline}
\Phi_{\text{TTRM}}(\theta, g_1, \alpha_1, g_2, \alpha_2, \gamma)
\\
=
\gamma\,\Phi_{\text{RM}}(\theta, g_1, \alpha_1) 
+ (1-\gamma)\,\Phi_{\text{RM}}(\theta, g_2, \alpha_2),
\end{multline}
where the RM phase function $\Phi_{\text{RM}}(\theta, g, \alpha)$ is given by 
\begin{multline}
\Phi_{\text{RM}}(\theta, g, \alpha)
\\=
\frac{\alpha}{\pi}\,\frac{g\,(1-g^2)^{2\alpha}}{(1+g)^{2\alpha}-(1-g)^{2\alpha}}\,
\frac{1}{(1+g^2-2g\cos\theta)^{1+\alpha}}.
\end{multline}
As the RM phase function itself is already a generalisation of the HG phase function, it is only natural that the five-parameter TTRM phase function can cover a much richer variety in phase function shapes. In particular, by choosing $g_1>0$ and $g_2<0$ it can describe phase functions with a distinctive forward and backward scattering peak. 

Again, we minimised the relative difference $h_{\text{rel}}$ to determine the parameters of the best-fitting TTRM phase function to the empirical phase functions at wavelengths between 0.1 and 5 $\mu$m. Not surprisingly, the result is a significantly improved fit. This fit is shown as the cyan lines in Figs.~{\ref{all-fits.fig}} and~{\ref{pp-polar.fig}}. The TTRM phase function reproduces the shape of the empirical phase functions very well at all wavelengths, and the relative difference $\delta_{\text{rel}}(\theta)$ remains under 5\% for all bands and all scattering angles. Only for the UV bands, $\delta_{\text{rel}}(\theta)$ slightly exceeds 5\% around the forward and the backward directions.

\section{The TTU2 phase function}
\label{newpf.sec}

\subsection{Towards a new three-parameter phase function}

\begin{figure}
\centering
\includegraphics[width=0.8\columnwidth]{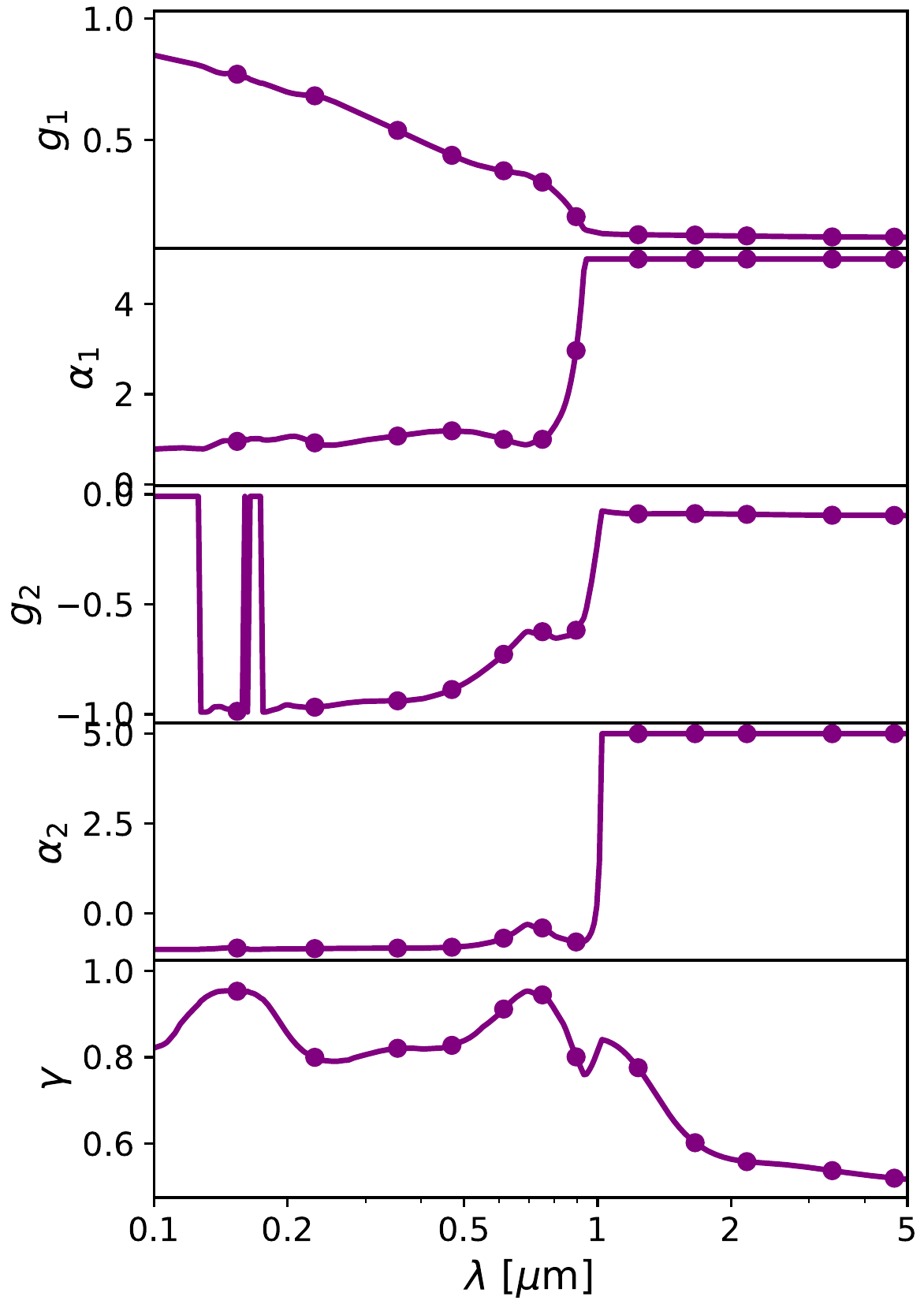}\hspace*{2em}
\caption{Model parameters of the best-fitting TTRM phase functions as a function of wavelength. The dots indicate the effective wavelengths of the GALEX, SDSS, 2MASS, and WISE broadband filters.} 
\label{TTRM-modelparameters.fig}
\end{figure}

In principle, our analysis could stop here: in line with \citet{C9NR01707K} and \citet{Harmel2021} we have reached the conclusion that the TTRM phase function provides a superior fit to the scattering phase function of interstellar dust. However, this phase function contains five free parameters, which makes it rather complex. Moreover, a closer analysis of the fitting results shows some interesting results. 

Fig.~{\ref{TTRM-modelparameters.fig}} shows the model parameters of the best-fitting TTRM phase function as a function of wavelength. This figure shows a number of peculiarities of the fits. Throughout the UV and optical range, the parameter $\alpha_1$ tends to prefer values around $\alpha_1=1$, whereas the parameter $\alpha_2$ appears to prefer negative values, and usually the lowest possible value $\alpha_2=-1$. For $\alpha=-1$, the RM phase function reduces to the isotropic phase function $\Phi_{\text{iso}}(\theta)=1/4\pi$, irrespective of the value of $g$. This makes the value of $g_2$ at UV and optical values largely irrelevant, which explains the spurious jumps of $g_2$ between $-1$ and 0 in the third panel of Fig.~{\ref{TTRM-modelparameters.fig}}. In summary, at UV and optical wavelengths, the phase function is well fitted by a linear combination of a forward scattering phase function with $\alpha_1\sim1$ and an isotropic phase function. Around $\lambda=1$~$\mu$m, both values of $\alpha$ shoot to $\alpha_{\text{max}} = 5$, the upper bound for $\alpha$ allowed by our fitting routine, while the $g$ parameters converge to $g_1=-g_2 = 0.098$. Interestingly, when the upper bound for $\alpha$ in the fitting routine is changed, the same qualitative behaviour persists. For example, with $\alpha_{\text{max}} = 10$, both $\alpha_1$ and $\alpha_2$ quickly rise to this value beyond $\lambda\sim1$~$\mu$m, while the $g$ parameters converge to $g_1=-g_2=0.056$. The difference in $h_{\text{rel}}$ between these models is negligible.

This strange behaviour clearly shows that the TTRM model contains a number of degeneracies, and suggests that a five-parameter phase function is probably overkill to describe the scattering phase function. Guided by the fact that the best-fitting value of $\alpha_1$ remains rather constant over the entire UV and optical range, and that the quality of the fits in the NIR seem to be relatively insensitive to the value of $\alpha_1$ and $\alpha_2$, we propose a restricted TTRM phase function, in which we fix $\alpha_1 = \alpha_2$ to a single fixed value $\alpha$ that is a wavelength-independent model parameter. The obvious choice would be $\alpha = \tfrac12$, in which case we end up with the TTHG phase function. The fact that, for our general five-parameter TTRM fit, $\alpha_1$ always tends to prefer higher values than $\tfrac12$ suggests that this might not the optimal choice. 

\begin{figure}
\centering
\includegraphics[width=0.9\columnwidth]{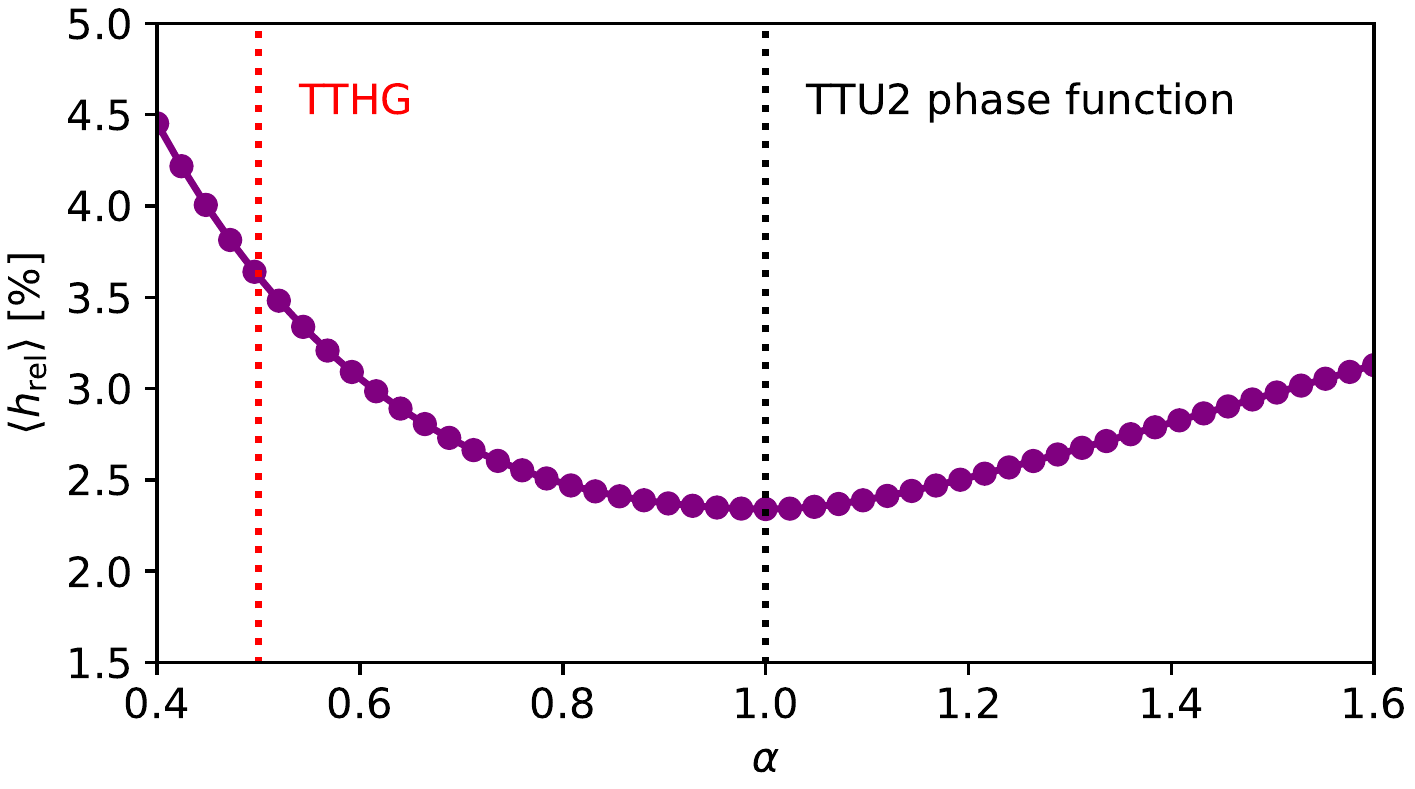}\hspace*{2em}
\caption{Relative error $\langle h_{\text{rel}}\rangle$, averaged over the standard UV, optical, and NIR broadband filters, for the TTRM phase function as a function of the fixed $\alpha=\alpha_1=\alpha_2$.} 
\label{bestalpha.fig}
\end{figure}

\begin{figure*}
\centering
\includegraphics[width=0.7\textwidth]{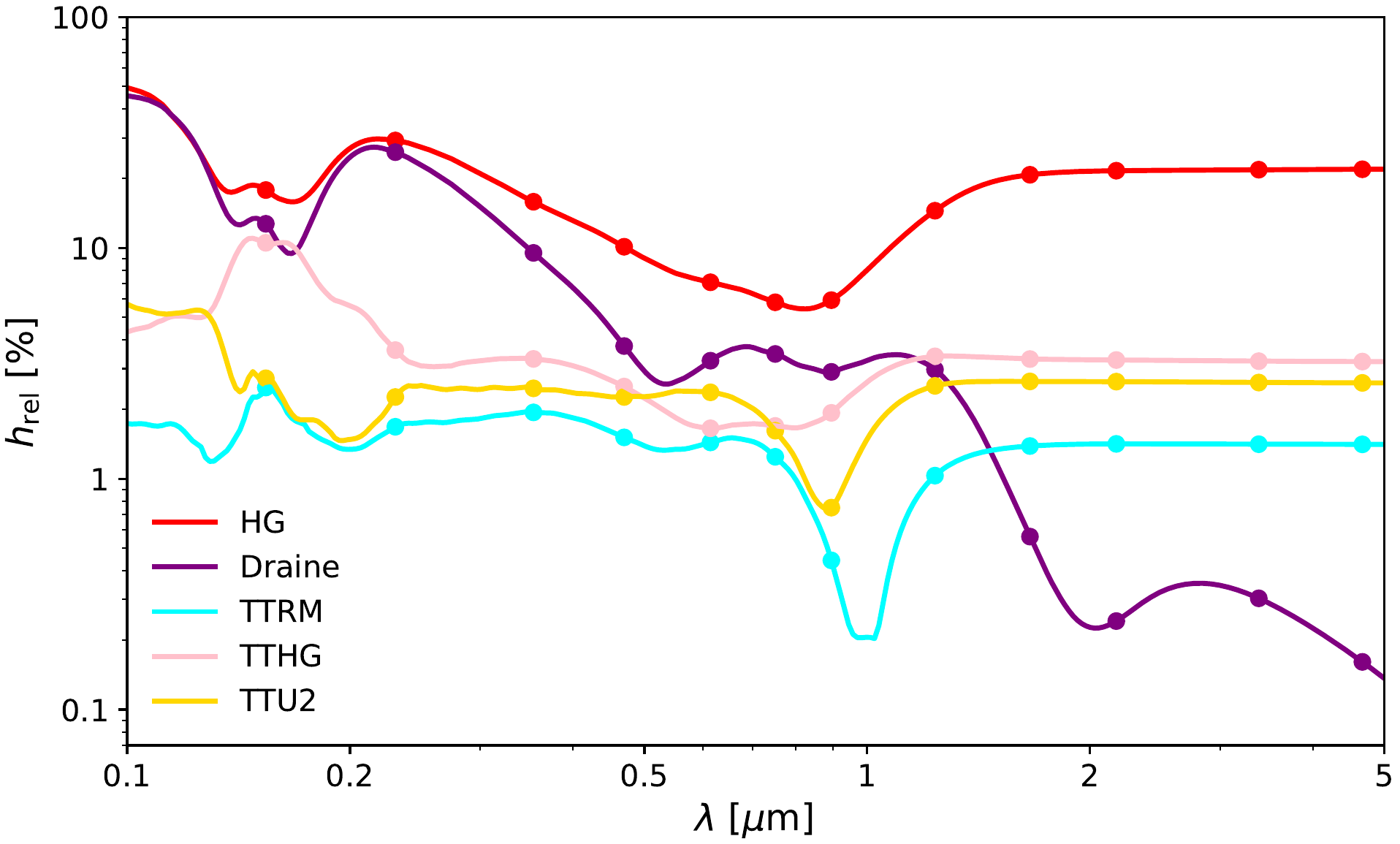}\hspace*{2em}
\caption{Comparison of the relative error $h_{\text{rel}}$ as a function of wavelength for the different phase functions considered in this paper. The dots indicate the effective wavelengths of the GALEX, SDSS, 2MASS, and WISE broadband filters.} 
\label{hrel.fig}
\end{figure*}

We therefore repeated our fitting exercise using different values of $\alpha$. At every wavelength, we fitted a constrained TTRM phase function with $\alpha_1=\alpha_2=\alpha$ to the empirical data, and determined the best-fitting values of the four free parameters ${\boldsymbol{p}} = (g_1,g_2,\alpha,\gamma)$. We finally determined the best value of $\alpha$ as the one that has the lowest relative error between the model fits and the empirical phase function, averaged over 12 wavelengths in the $0.1-5~\mu$m range.\footnote{We consider the effective wavelengths of the GALEX FUV and NUV bands, the five optical SDSS bands, the 2MASS {\em{J}}, {\em{H,}} and {\em{K}}$_{\text{s}}$ bands, and the WISE {\it{W1}} and {\it{W2}} bands.} Fig.~{\ref{bestalpha.fig}} shows this average relative error $\langle h_{\text{rel}} \rangle$ as a function of $\alpha$. It turns out that the optimal value for $\alpha$ is indeed not $\tfrac12$, the value corresponding to the TTHG phase function. Interestingly, the value for $\alpha$ that minimises $\langle h_{\text{rel}} \rangle$ is found to be exactly equal to one. 

\subsection{Characteristics of the TTU2 phase function}

Putting all pieces together, we find the following expression for a new analytical phase function,
\begin{equation}
\Phi_{\text{TTU2}}(\theta,g_1,g_2,\gamma)
= \gamma\,\Phi_{\text{U2}}(\theta, g_1)+ (1-\gamma)\,\Phi_{\text{U2}}(\theta, g_2)
\label{TTU2}
,\end{equation}
where 
\begin{equation}
\Phi_{\text{U2}}(\theta, g)
=
\frac{1}{4\pi}\,\frac{(1-g^2)^2}{(1+g^2-2g\cos\theta)^2}.
\label{U2}
\end{equation}
We propose the name ultraspherical-2 (U2) phase function for the simple phase function~(\ref{U2}), since it is the generating function for the set of ultraspherical polynomials of order 2 (Appendix~{\ref{Gegenbauer.sec}}). In analogy with the TTRM and TTHG phase functions, we then propose the name two-term ultraspherical-2 (TTU2) phase function for the phase function represented by Eq.~(\ref{TTU2}). It is easy to verify that the U2 phase function, and therefore also the TTU2 phase function, satisfies the normalisation convention~(\ref{norm}).

The TTU2 phase function is characterised by three free parameters with the following meanings. First, the parameter $g_1$, with values between 0 and 1, characterises the shape and the strength of the forward scattering peaking part of the phase function. The larger the value of $g_1$, the more asymmetric and peaked the forward scattering part of the phase function. If $g_1=0$, the forward scattering part is isotropic, if $g_1=1$, it is completely forward scattering. Similarly, $g_2$, with values between --1 and 0, characterises the shape and strength of the backward scattering peak. Third, the parameter $\gamma$ sets the relative weight of both components and can take any value, as long as the phase function remains positive for all scattering angles. We limit its range to $0\leqslant\gamma\leqslant1$. 

It is important to realise that the parameter $g$ that characterises the U2 phase function is not the anisotropy parameter, contrary to the case of the HG phase function. For the U2 phase function we find instead
\begin{equation}
\langle\cos\theta\rangle_{\text{U2}} = \frac{1+g^2}{2g}+\left(\frac{1-g^2}{2g}\right)^2\ln\left(\frac{1-g}{1+g}\right).
\label{mcos}
\end{equation}
For the TTU2 three-parameter phase function (\ref{TTU2}), expression~(\ref{mcos}) is readily generalised to
\begin{multline}
\label{asymmpar}
\langle\cos\theta\rangle_{\text{TTU2}}  =
\gamma \left[\frac{1+g_1^2}{2g_1}+\left(\frac{1-g_1^2}{2g_1}\right)^2\ln\left(\frac{1-g_1}{1+g_1}\right)\right]
\\
+(1-\gamma) \left[\frac{1+g_2^2}{2g_2}+\left(\frac{1-g_2^2}{2g_2}\right)^2\ln\left(\frac{1-g_2}{1+g_2}\right)\right].
\end{multline}

\subsection{Applicability to interstellar dust}

The fits of the TTU2 phase function to the empirical BARE-GR-S phase functions are shown as the golden lines in Fig.~{\ref{all-fits.fig}}. Compared to the general TTRM phase function, the fits are hardly poorer, in spite of the fact that we only fit three instead of five free parameters. Especially in the UV and the optical, the TTU2 phase function provides an almost equally good fit to the empirical data, with almost identical relative differences. At NIR wavelengths the TTU2 phase function slightly overestimates the forward and backward scattering peak and also the scattering around the minimum at 90$^\circ$. The relative error remains within the 5\% range, however.

As a summary, Fig.~{\ref{hrel.fig}} compares the relative error $h_{\text{rel}}$ of the different phase functions discussed as a function of wavelength over the wavelength range from 0.1 to 5~$\mu$m. The HG phase function, which only contains one free parameter, clearly provides the poorest fit to the empirical phase function over the entire wavelength range. Relative errors range between $\gtrsim$10\% at 1~$\mu$m to almost 100\% at the shortest wavelengths. At UV and optical wavelengths, the two-parameter Draine phase function performs only marginally better than the HG phase function, but at NIR wavelengths, this phase function outperforms all other options. The other three phase functions shown are all two-term phase functions. The TTRM phase function, with its five free parameters, obviously has the best performance, with a relative error of a few percent over the entire wavelength range. The new TTU2 phase function, with only three free parameters, is only slightly poorer. It is characterised by a relative error below 5\% over the entire optical and NIR wavelength range, and only slightly larger at the shortest wavelengths. The TTHG phase function, also with three free parameters, consistently performs more poorly than the TTU2 phase function over the entire wavelength range considered, except in a small wavelength interval between 0.6 and 0.74~$\mu$m and at the shortest UV wavelengths (below 0.12~$\mu$m). Based on this plot and the previous discussion, we argue that the TTU2 phase function is the compromise looked for: it provides an accurate fit to the empirical phase function over a wide wavelength range, and with three free parameters it is still simple enough.

\begin{figure}
\centering
\includegraphics[width=0.9\columnwidth]{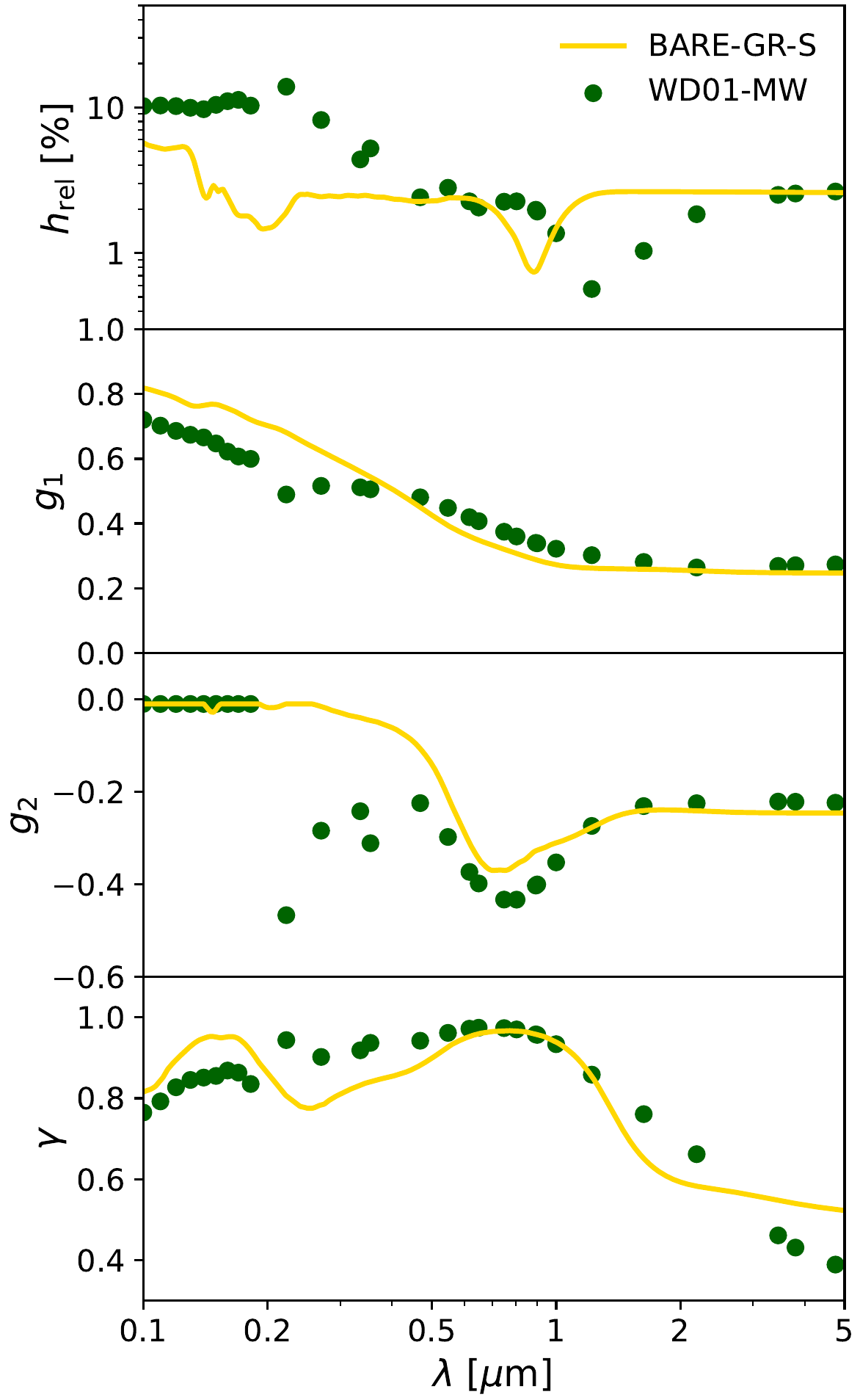}\hspace*{2em}
\caption{Relative error and model parameters of the best-fitting TTU2 phase functions as a function of wavelength. The golden lines correspond to our default BARE-GR-S dust model, and the green dots correspond to the alternative WD01-MW dust model.} 
\label{TTU2-modelparameters-all.fig}
\end{figure}

\begin{table}
\begin{tabular}{ccccc}
band & $\lambda$ & $g_1$ & $g_2$ & $\gamma$ \\
& [$\mu$m] & & & \\[1mm]
 \hline \\[-2mm]
FUV & $0.154$ & $0.764$ & $-0.016$ & $0.950$ \\
NUV & $0.230$ & $0.670$ & $-0.010$ & $0.792$ \\
{\em{u}} & $0.356$ & $0.545$ & $-0.048$ & $0.841$ \\
{\em{g}} & $0.470$ & $0.449$ & $-0.109$ & $0.882$ \\
{\em{r}} & $0.618$ & $0.361$ & $-0.308$ & $0.952$ \\
{\em{i}} & $0.749$ & $0.320$ & $-0.370$ & $0.967$ \\
{\em{z}} & $0.895$ & $0.288$ & $-0.327$ & $0.960$ \\
{\em{J}} & $1.239$ & $0.262$ & $-0.274$ & $0.846$ \\
{\em{H}} & $1.649$ & $0.259$ & $-0.240$ & $0.643$ \\
{\em{K}}$_{\text{s}}$ & $2.164$ & $0.254$ & $-0.241$ & $0.583$ \\
{\em{W1}} & $3.390$ & $0.248$ & $-0.245$ & $0.549$ \\
{\em{W2}} & $4.641$ & $0.247$ & $-0.246$ & $0.526$ \\
\end{tabular}
\caption{Model parameters of the best-fitting TTU2 phase functions for the BARE-GR-S dust model at the central wavelengths of a number of standard broadband filters. Files with the model parameters for the BARE-GR-S and WD01-MW dust models are available in Appendix~{\ref{TTU2-paramfiles.sec}}.}
\label{TTU2-params.tab}
\end{table}

In order to better understand the meaning of the three parameters of the TTU2 phase function, we show them explicitly as a function of wavelength in Fig.~{\ref{TTU2-modelparameters-all.fig}}, and we list the values at the central wavelengths of a number of standard broadband filters in Table~{\ref{TTU2-params.tab}}. The bottom panel of Fig.~{\ref{TTU2-modelparameters-all.fig}} shows that the forward scattering component of the TTU2 phase function always dominates over the backward scattering component. Interestingly, the parameter $\gamma$ does not decrease monotonically as a function of wavelength: it first rises to a maximum of more than 0.9 at the FUV band, subsequently falls back to slightly below 0.8 in the NUV band, and then gradually increases to 0.9 again around 0.8~$\mu$m. In the NIR the importance of the first component gradually decreases towards the asymptotic value of 0.5 expected for a Rayleigh scattering phase function. This behaviour needs to be interpreted in conjunction with the parameters $g_1$ and $g_2$. At UV wavelengths, the forward scattering component is strongly peaked, whereas the backward scattering component is essentially isotropic. Moving towards optical wavelengths, the value of $g_1$ gradually drops, whereas the backward component becomes more anisotropic. In the NIR regime, both components tend towards the same (opposite) value of moderate anisotropy, $g_1\approx-g_2\approx0.246$.

We repeated our analysis for an alternative dust model: the WD01-MW dust model, that is, the average Milky Way dust mixture presented by \citet{2001ApJ...548..296W}. This model also consists of a mixture of graphite, silicates, and PAHs, but it is based on different optical properties and a different size distribution. Within our wavelength range from 0.1 to 5 $\mu$m, explicit scattering phase functions with a 1~deg resolution are available for 27 different wavelengths on Bruce Draine's web page\footnote{\url{https://www.astro.princeton.edu/~draine/}}. Fitting the different analytical phase functions to the empirical WD01-MW phase functions, we find a similar qualitative behaviour as for our default BARE-GR-S model. The one-parameter HG model generally results in the poorest fit (relative differences of several ten percent), the five-parameter TTRM model in the best fit (relative differences of a few percent), and the three-parameter TTU2 model in an almost equally good fit. The top panel of Fig.~{\ref{TTU2-modelparameters-all.fig}} shows that the TTU2 phase function provides an excellent fit to the phase function of the WD01-MW dust mixture in the optical and NIR range, with relative errors of the order of a few percent. In the UV part of the spectrum, the fits are slightly poorer, with relative errors around 10\%. 

The remaining three panels of Fig.~{\ref{TTU2-modelparameters-all.fig}} show that, in spite of the differences between the BARE-GR-S and WD01-MW models, the fitted parameters of the TTU2 phase function agree fairly well. For the WD01-MW dust model too the prime component gradually becomes less forward scattering when moving to longer wavelengths, the second component is isotropic in the UV and more strongly anisotropic at optical wavelengths, and $\gamma$ peaks at $\lambda\sim 0.8~\mu{\text{m}}$. The main difference between the two models is a spurious jump for the different parameters, especially $g_2$, between 0.17 and 0.182~$\mu$m. This apparent discontinuity reflects a sudden transition that is primarily seen in the backward scattering peak of the WD01-MW phase function.

\subsection{Integration in radiative transfer codes}

As stated in the Introduction, our objective was to search for analytical scattering phase function with two competing characteristics. On the one hand it should be flexible enough to accurately describe the actual scattering phase function for interstellar dust over the UV--NIR wavelength range, while on the other hand it should be simple enough to be easy to use, especially in the context of radiative transfer calculations. 

For 1D radiative transfer problems in either plane-parallel or spherical geometry, there are various options to solve the radiative transfer equation. One approach is the so-called spherical harmonics or $P_L$ method, which builds on the expansion of the radiation field as a series of Legendre polynomials. This turns the dust radiative transfer equation, in general an integro-differential equation, into an infinite set of ordinary differential equations. This set is turned into a finite set of equations by imposing a closure condition, that is, by truncating the Legendre polynomial expansion. This set of equations can be reformulated as a classic eigenvalue problem and thus solved by standard matrix diagonalisation. Detailed descriptions of this method can be found in \citet{1980ApJ...236..598F}, \citet{1983ApJ...275..292R} and \citet{1995MNRAS.277.1279D}. In \citet{2001MNRAS.326..722B} we compared four different methods to solve the dust radiative transfer methods in a plane-parallel geometry, and we found that the spherical harmonics method was by far the most efficient one. Furthermore, \citet{2018ApJ...861...80C} and \citet{2021A&A...645A.143K} demonstrated that the spherical harmonics method can be applied in optical depth regimes not easily accessible to Monte Carlo codes. 

In most astrophysical contexts, the geometry of the objects under study is complex on different scales, which makes 3D radiative transfer a necessity. The Monte Carlo technique, which uses a probabilistic approach to simulate the individual life cycles of a large number of individual photon packets, is currently by far the most widely used technique for 3D dust radiative transfer. There are several codes available that can solve the dust radiative transfer problem in an arbitrary 3D setting \citep[e.g.][]{2001ApJ...551..269G, 2011A&A...536A..79R, 2015A&C.....9...20C, 2020A&C....3100381C, 2016A&A...593A..87R, 2019A&A...622A..79J}. Most codes that participated in the TRUST 3D dust radiative transfer benchmark effort \citep{2017A&A...603A.114G} were based on the Monte Carlo method. For an overview of Monte Carlo radiative transfer we refer to \citet{2011BASI...39..101W}, \citet{2013ARA&A..51...63S}, and \citet{2019LRCA....5....1N}. 

In the following two subsections we discuss whether the new TTU2 phase function is easily integrated into the spherical harmonics and Monte Carlo radiative transfer frameworks. We particularly investigate whether the transition from the HG to the TTU2 phase function involves a significant increase in complexity or overhead.

\subsubsection{Spherical harmonics radiative transfer}

As mentioned above, the bottom-line of the spherical harmonics method is that the radiation field is expanded in a series of Legendre polynomials. The scattering phase function enters the spherical harmonics methods by means of its expansion in Legendre polynomials,
\begin{equation}
\Phi(\theta) = \frac{1}{4\pi} \sum_{\ell=0}^\infty (2\ell+1)\,\sigma_\ell\,P_\ell(\cos\theta),
\end{equation}
with the coefficients $\sigma_\ell$ given by
\begin{equation}
\sigma_\ell = \int \Phi(\theta)\,P_\ell(\cos\theta)\,{\text{d}}\Omega.
\label{sigmaell}
\end{equation}
The set of coefficients $\{\sigma_\ell\}_{\ell\leqslant M}$, with $M$ the chosen order of the Legendre polynomial series truncation, fully characterises the scattering phase function in this method.

The HG phase function is extremely well suited for the spherical harmonics method since its expansion in Legendre polynomials is very simple \citep{1983ApJ...275..292R}:
\begin{equation}
\sigma_\ell = g^\ell.
\end{equation}
If we want to replace this phase function with the TTU2 phase function, we have to calculate the coefficients $\sigma_\ell$ by inserting the expression (\ref{TTU2}) in the integral~(\ref{sigmaell}). The resulting integral can be evaluated analytically for each value of $\ell$. For the first two coefficients we obviously find 
\begin{gather}
\sigma_0 = 1, \\
\sigma_1 = \langle\cos\theta\rangle_{\text{TTU2}} .
\end{gather}
For higher values of $\ell$, the resulting expressions are similar to expression (\ref{asymmpar}), but with higher-order polynomials in each of the terms. While this is clearly less attractive than the simple expression for the HG phase function, this is not a major source of overhead or increased complexity: the calculation of the Legendre coefficients $\sigma_\ell$ can be performed numerically at the beginning of the actual radiative transfer calculations, or even beforehand.

An alternative, exploratory, option could be to use the fact that the U2 phase function, and hence also the TTU2 phase function, has a simple expansion in terms of ultraspherical polynomials of order 2 (see Appendix~{\ref{Gegenbauer.sec}}),
\begin{equation}
\Phi_{\text{U2}}(\theta;g)
=
\frac{(1-g^2)^2}{4\pi}
\sum_{n=0}^\infty
g^n\,C_n^{(2)}(\cos\theta).
\end{equation}
The conventional spherical harmonics method builds on an expansion of the radiation field in terms of Legendre polynomials, but this is not the only option. Another option is an expansion in terms of Chebyshev polynomials of the first or second kind, which might result in better approximate solutions \citep[e.g.][]{Conkie1959, Milgram1991, Yasa2006, 2006JQSRT.101..129A, 2007JQSRT.105..211O, Bulbul2008}. Legendre and Chebyshev polynomials are special cases of the family of ultraspherical polynomials, and it is in principle also possible to expand the radiation field in ultraspherical polynomials of order 2. Some efforts towards a general expansion in ultraspherical polynomials have been undertaken in the context of neutron and heat transport \citep{Yilmazer2008a, Yilmazer2008b}. It remains to be seen how complex the extension towards radiative transfer with a general anisotropic phase functions will turn out to be; this is clearly beyond the scope of this paper.

\subsubsection{Monte Carlo radiative transfer}

\begin{figure}
\centering
\includegraphics[width=0.9\columnwidth]{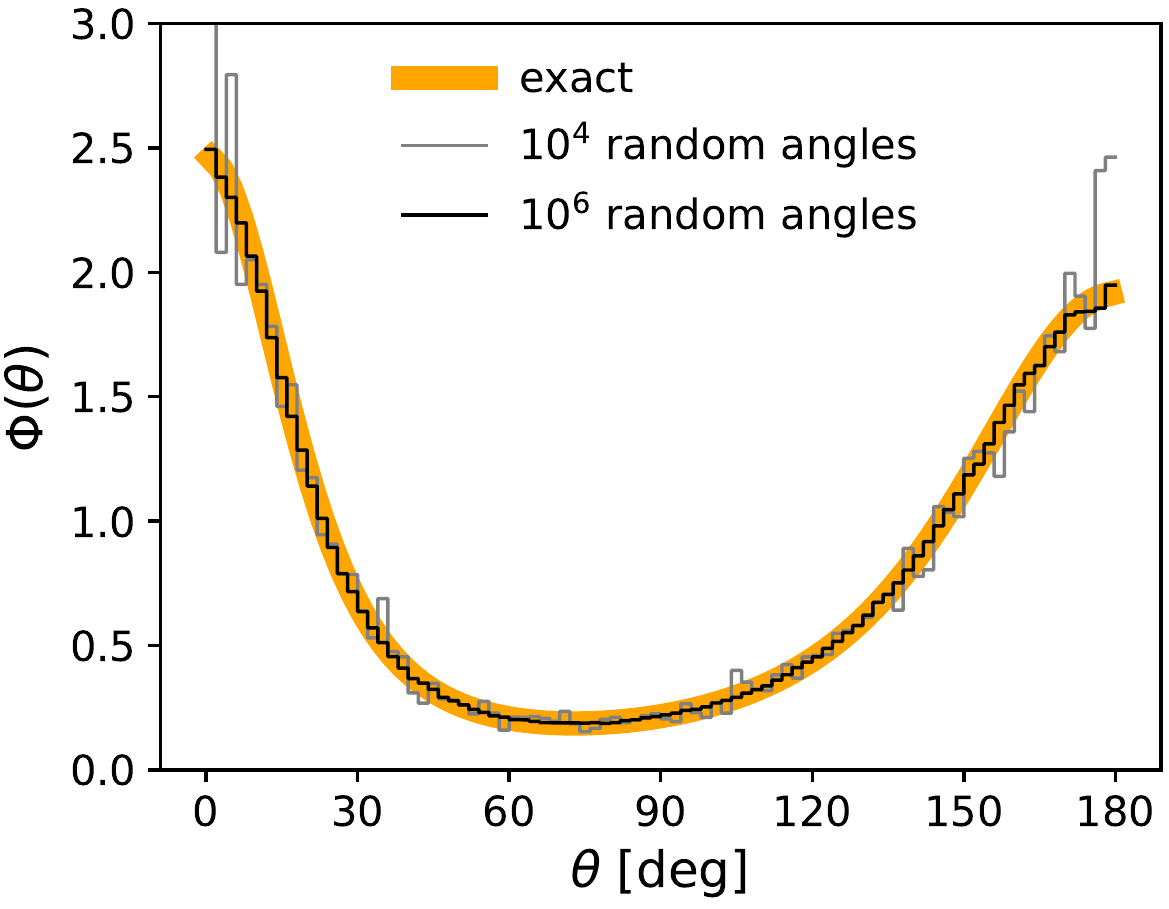}\hspace*{2em}
\caption{Demonstration of the method to extract random scattering angles from the TTU2 phase function. The solid orange line corresponds to the TTU2 phase function with $g_1=0.6$, $g_2=-0.4$ and $\gamma=0.3$. The grey and black lines are reconstructions of the phase function based on randomly generated scattering angles.} 
\label{TTU2-random.fig}
\end{figure}

In Monte Carlo radiative transfer, a new scattering angle $\theta$ has to be randomly generated from the scattering phase function after every single scattering event. As every simulation typically requires millions or billions of scattering events, it is crucial that this operation is efficient and accurate. The standard way to do this is by means of inversion sampling \citep{devroye2013non}: one chooses a random deviate $X$ and solves the equation $X = P(\theta)$, where $P(\theta)$ is the cumulative distribution function of the phase function
\begin{equation}
P(\theta) = 2\pi \int_0^\theta \Phi(\theta') \sin\theta'\, {\text{d}}\theta'.
\end{equation}
For the HG phase function, the cumulative distribution function can be calculated analytically and the resulting expression can be inverted exactly \citep{1977ApJS...35....1W},
%\begin{equation}
%P(\theta) = \frac{1-g^2}{2g}\left(\frac{1}{\sqrt{1+g^2-2g\cos\theta}}-\frac{1}{1+g}\right).
%\end{equation}
\begin{equation}
\theta = \arccos\left\{\frac{1}{2g}\left[1+g^2 - \left(\frac{1-g^2}{1-g + 2gX}\right)^2\right]\right\}.
\label{muHG}
\end{equation}
%This expression can be inverted analytically, such that a new scattering angle $\theta$ is obtained as
This makes the inclusion of the HG phase function in the Monte Carlo framework very attractive.

For the U2 phase function we find that the cumulative distribution function of scattering angles can also be expressed analytically,
\begin{equation}
P(\theta) = \frac{(1+g)^2\,(1-\cos\theta)}{2\,(1+g^2-2g\cos\theta)}.
\label{cumU2}
\end{equation}
This expression can also be inverted analytically, resulting in the following recipe to generate a random scattering angle for the U2 phase function,
\begin{equation}
\theta = \arccos\left[\frac{(1+g)^2-2X(1+g^2)}{(1+g)^2-4gX}\right].
\label{ranthetaU2}
\end{equation}
So the random generation of scattering angles can also be done in a very straightforward way for the U2 phase function.

For the TTU2 phase function, we obviously get for the cumulative distribution function a linear combination of two terms of the form~(\ref{cumU2}), but, unfortunately, this expression cannot be inverted analytically anymore. This implies that we need other ways to generate random scattering angles. One option could be to use a numerical tabulation of the scattering phase function, as adopted by \citet{1977ApJS...35....1W} for his Monte Carlo radiative transfer calculations using the TTHG phase function. 

A more elegant, efficient and accurate approach consists of using the principle of decomposition in random number generation \citep{devroye2013non, 2015A&C....12...33B}. As a first step, we generate a first random uniform deviate $U$ and determine which of the two U2 components will be selected. The importance of both components is determined by their relative weight: if $0<U<\gamma$ the scattering angle will be generated from the first component, if $\gamma<U<1$ it is the second component. Subsequently, we generate a second uniform deviate $X$ and determine the scattering angle according to Eq.~(\ref{ranthetaU2}), where we replace $g$ by either $g_1$ or $g_2$, depending on the component selected in the first step. In fact, the procedure can be made even more efficient by reusing the first random deviate in the second step. Rather than generating a second uniform deviate for the second step, we can also set
\begin{equation}
X = \begin{cases}
\;U/\gamma & \quad 0<U<\gamma, \\
\;(1-U)/(1-\gamma) & \quad \gamma<U<1.
\end{cases}
\end{equation}
This procedure avoids the generation of a second uniform deviate and makes the algorithm to generate random scattering angles from the three-parameter TTU2 phase function essentially as efficient as for the HG phase function.

The accuracy of this algorithm is shown in Fig.~{\ref{TTU2-random.fig}}. The thick orange line in this figure corresponds to the exact TTU2 phase function with arbitrarily chosen parameters $g_1=0.6$, $g_2=-0.4$ and $\gamma=0.3$. The grey line corresponds to the phase function reconstructed from the histogram of $10^4$ randomly generated scattering angles, based on bin widths of 2~deg. The phase function is accurately reproduced, with some Poisson noise, especially around the forward and backward scattering peaks. The black line is similar to the grey line, but now corresponds to $10^6$ random scattering angles. The Poisson noise is now reduced significantly.

In summary, we find that the three-parameter TTU2 function can easily and efficiently be included in the Monte Carlo framework.

\section{Summary}
\label{Summary.sec}

The goal of this paper was to search for alternative analytical scattering phase functions to describe the scattering properties of dust grains in the interstellar medium of galaxies. We specifically aimed at finding a balance between realism and complexity: on the one hand the scattering phase function should be flexible enough to provide an accurate fit to the actual scattering properties of dust grains over a wide wavelength range, and on the other hand it should be simple enough to be easy to handle, especially in the context of radiative transfer calculations. 

We considered various analytical phase functions available in the literature and fitted them to `empirical' data corresponding to the BARE-GR-S model by presented by \citet{2004ApJS..152..211Z}. We considered the UV--NIR wavelength range, corresponding to wavelengths between 0.1 and 5~$\mu$m. Our findings are the following:
\begin{itemize}
\item The standard one-parameter HG phase function provides a good fit to the empirical data at optical wavelengths but fails at describing the forward and backward scattering peak at UV and NIR wavelengths. Relative differences between the empirical data and the best-fitting HG phase function reach 30\% in the NIR and 50\% in the UV.
\item The two-parameter Draine phase function introduced by \citet{2003ApJ...598.1017D} was designed as a generalisation of both the HG phase function and the Rayleigh phase function. It provides an excellent fit at NIR wavelengths but fails to alleviate the discrepancies in the UV.
\item The TTRM phase function was recently advocated by \citet{C9NR01707K} and \citet{Harmel2021} in the context of light scattering in nanoscale materials and aquatic media, respectively. We find that it provides an excellent fit to the empirical BARE-GR-S phase function data over the entire UV-NIR range. We do, however, find degeneracies between its five free parameters, suggesting that a five-parameter phase function is overkill for describing the scattering phase function. 
\item We propose a simpler phase function, the two-term ultraspherical-2 (TTU2) phase function, that depends on only three free parameters. We show that it provides almost equally good fits to the empirical data as the TTRM phase function over the entire wavelength range considered, with relative differences almost always below 5\%.
\end{itemize}
The new TTU2 phase function is characterised by three free parameters with a simple physical interpretation: two parameters characterise the shape of two components of the phase function, a forward and a backward scattering one, and the third parameter sets the relative weight of both components. We have investigated how these parameters change with wavelength, both for the BARE-GR-S model and an alternative dust model. In spite of the differences between the two dust models, the results are compatible:
\begin{itemize}
\item The forward scattering component of the phase function systematically becomes less anisotropic as the wavelength increases. 
\item The backward scattering component is isotropic in the UV and moderately anisotropic at optical and NIR wavelengths. The anisotropy peaks in the optical regime. 
\item The relative strength of the backward scattering component peak is of the order of 20\% at UV and blue wavelengths, has a minimal value of about 3\% in the red ($\lambda\sim0.8~\mu$m), and gradually increases towards 50\% at longer wavelengths. 
\end{itemize}
Finally, we have investigated whether the integration of the TTU2 phase function in radiative transfer calculations implies a significant overhead or increased complexity. We have specifically focused on the spherical harmonics method and the Monte Carlo method: the former is one of the most efficient methods for 1D radiative transfer, and the latter is the most popular option for 3D radiative transfer. While the standard HG phase function is easily and almost naturally integrated into both approaches, we find that the shift from the HG to the TTU2 phase function does not imply a major source of overhead or increased complexity. 

Our overall conclusion is that the TTU2 scattering phase function that we have proposed in this paper provides a nice balance between being simple enough to be easily used and realistic enough to accurately describe scattering by dust grains. We advocate its employment in astrophysical applications, in particular in radiative transfer calculations. An investigation of the effects of the choice of the phase function in radiative transfer calculations in different optical depth regimes \citep[see e.g.][]{2001MNRAS.326..733B} will be considered in future work.

\begin{acknowledgements}
The authors thank the anonymous referee for insightful and detailed comments that improved the content and presentation of this work. This paper is dedicated to the memory of Jonathan I. Davies (1955--2021). He was an inspiring researcher, mentor, leader, colleague and friend, and will be deeply missed.
\end{acknowledgements}

\bibliography{TTU2}

\appendix
\section{Ultraspherical polynomials}
\label{Gegenbauer.sec}

The ultraspherical or Gegenbauer polynomials $C_n^{(\lambda)}(x)$ are orthogonal polynomials on the interval $[-1,1]$ with respect to the weight function $w(x) = (1-x^2)^{\lambda-1/2}$. Their explicit form can be obtained in different ways \citep{10.2307/2031975, Kim2012, Doman2015}. One of the most convenient ways is by means of the recurrence relation 
\begin{gather}
C_0^{(\lambda)}(x) = 1, \\
C_1^{(\lambda)}(x) = 2\lambda\,x, \\
C_n^{(\lambda)}(x) = \frac{2\,(n+\lambda-1)}{n}\,x\,C_{n-1}^{(\lambda)}(x) - \frac{n+2\lambda-2}{n}\,C_{n-2}^{(\lambda)}(x).
\end{gather}
Particularly important special cases of ultraspherical polynomials are obtained for particular values of the order $\lambda$: $\lambda=\tfrac12$ gives the Legendre polynomials $P_n(x)$, $\lambda=1$ gives the Chebyshev polynomials of the second kind $U_n(x)$, and $\lambda=0$ gives, in a suitable limiting form, the Chebyshev polynomials of the first kind $T_n(x)$.

The generating function for the ultraspherical polynomials is a simple rational function \citep{10.2307/2031975},
\begin{equation}
\frac{1}{(1-2xt+t^2)^\lambda} = \sum_{n=0}^\infty t^n\,C_n^{(\lambda)}(x).
\end{equation}
Comparing this expression to the U2 phase function~(\ref{U2}) we immediately find
\begin{equation}
\Phi_{\text{U2}}(\theta;g)
=
\frac{(1-g^2)^2}{4\pi}
\sum_{n=0}^\infty
g^n\,C_n^{(2)}(\cos\theta).
\end{equation}
The lowest-order ultraspherical polynomials of order 2 are
\begin{gather}
C_0^{(2)}(x) = 1, \\
C_1^{(2)}(x) = 4x, \\
C_2^{(2)}(x) = 12x^2-2, \\
C_3^{(2)}(x) = 32x^3-12x, \\
C_4^{(2)}(x) = 80x^4-48x^2+3.
\end{gather}
Due to their importance in spectral methods for the numerical solution of differential equations and computational fluid dynamics \citep{doi:10.1146/annurev.fl.19.010187.002011}, efficient methods for the calculation of the coefficients of ultraspherical polynomial expansions have been developed \citep{DEMICHELI2013112}.

\section{TTU2 phase function parameters}
\label{TTU2-paramfiles.sec}

The data files {\tt{BARE-GS-S-TTU2.txt}} and {\tt{WD01-MW-TTU2.txt}} contain the model parameters of the best-fitting TTU2 phase function to the empirical phase function for the BARE-GR-S and WD01-MW dust models, respectively. The former file contains fitting parameters for 1201 different wavelengths between 1~nm and 10~mm, the latter file for 36 different wavelengths between 60~nm and 10.2~$\mu$m.

\end{document}